\documentclass[b5paper,twoside]{jpconf}  
\usepackage{graphicx}

\begin{document}

\title[Search for and Characterization of Open Clusters with 2MASS]
{Search for and Characterization of Galactic Open Clusters with 2MASS}

\author[Lin \& Chen]{C. C. Lin$^{1,2}$, W. P. Chen$^1$, and E. A. Magnier$^{2}$}

\address{$^1$Graduate Institute Astronomy, National Central University, Taiwan}
\address{$^2$Institute for Astronomy, University of Hawaii, U.S.A}

\ead{m949006@astro.ncu.edu.tw}

\begin{abstract}
We have developed a star-counting algorithm and tested it on the 2MASS star catalog 
to search for density enhancements significantly above the field in Galactic latitude 
$|b|<50^{\circ}$.  Nearly 500 open clusters are ``rediscovered'', along with 89 globular 
clusters, 35 galaxies, 55 galaxy clusters, 11 H~II regions, and 4 regions contaminated by 
nearby bright stars.  Fifty-two density enhancement regions remain unaccounted for.  
Here we present one such candidate HDG\,01, $(\ell, b)=(144.9038, 0.4338)$ which 
has an angular size of $3'$, a distance of 1.5~kpc, hence a physical size of 3.7~pc. 
Due to some nebulous shape, this star cluster should be an young-aged ($\le 10$~Myr).
\end{abstract}

\section{Introduction}
It is estimated that some $10^{5}$ Galactic open clusters (OCs) should exist in the Milky Way 
Galaxy \cite{pis06}.  However, recent OC catalogs \cite{dnb01, dia02, bic03, 
dut03, kro06, fro07} list only a few thousands entries, mainly within 1~kpc.  
The discrepancy is due partly to the dust extinction in the 
Galactic plane, and partly to lack of comprehensive all-sky searches.

``Star counting'' is an efficient algorithm to identify star density enhancement on a wide-field 
or an all-sky survey \cite{sch11}.  Several existing all-sky survey databases can be used to 
identify star density enhancements as OC candidates.  For instance, the 2MASS point-source catalog \cite{skr06} 
provides a uniformly calibrated database of the entire sky.  Furthermore, infrared observations 
allow us to recognize OCs even with moderate dust extinction, i.e., partially embedded, young star 
clusters.  Recent work by \cite{bic03}, \cite{dut03} and \cite{fro07} have indeed found hundreds of 
previously unknown infrared clusters with 2MASS, some of which turn out to be bona fide stellar 
groups as verified by follow-up photometric studies.  Here we report the results of our searching
 algorithm and preliminary characterization of star clusters from 2MASS data.

\section{Data Analysis: Star Couning Algorithm}
The analyzed sky coverage was between Galactic latitude $|b|<50^{\circ}$.  We divided 
the sky into $2^{\circ}\times2^{\circ}$ fields and selected 2MASS stars with S/N$>5$ in all 
JHKs bands.  Each field was analyzed by counting the number of stars in a grid, whose size 
was chosen to include an average of 10 stars.  The same field then would be analyzed again
by shifting half a grid size.  The resulting density distribution was then smoothed with 
a $3 \times 3$ box.  The background density was estimated by $3\sigma$ clipping 
of outliers.  Any grid in the smoothed density map above $5\sigma$ background was 
considered to be a high density grid and a group of more than three adjacent high density 
grids would be identified as a high density group (HDG).  Our analysis resulted in a total 
of 720 HDGs.

\begin{figure}[h]
  \centering
  \includegraphics[width=25pc]{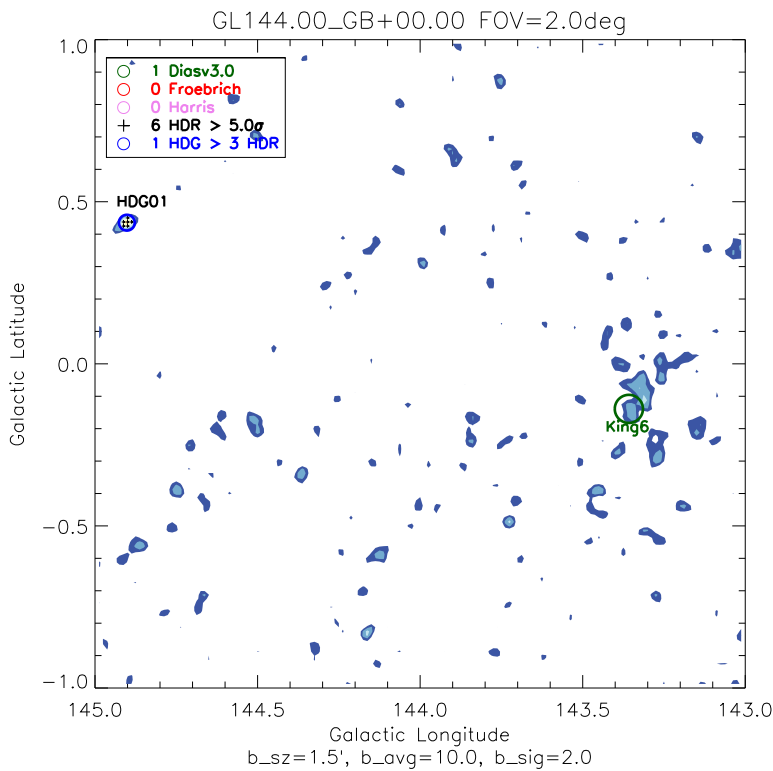}
  \caption{\label{sample}Demonstration of a high density group, HDG\,01.  The contour marks the  
  the 3$\sigma$ level above the sky density.  Known OCs (in green circles) are from \cite{dia02} version 3.0.  
  Note that \cite{fro07} and \cite{har96} catalogs were not found in this FOV.  Each black plus represents a high 
  density grid and a blue circle marks a HDG.}
\end{figure}

Using the SIMBAD database, we matched the HDGs within a radius of $5'$ with 668 known objects, 
including 474 OCs, 89 globular clusters, 35 galaxies, 55 galaxy clusters, 11 H~II regions, and 
4 regions contaminated by nearby bright stars.  A total of 52 HDGs remain unaccounted for.  
These are our OC candidates.  In Figure~\ref{sample}, we demonstrate such a case, HDG\,01 at 
$(\ell, b)=(144.9038, 0.4338)$, located near the Galactic plane.  

\begin{figure*}[h]
\centering\includegraphics[width=30pc]{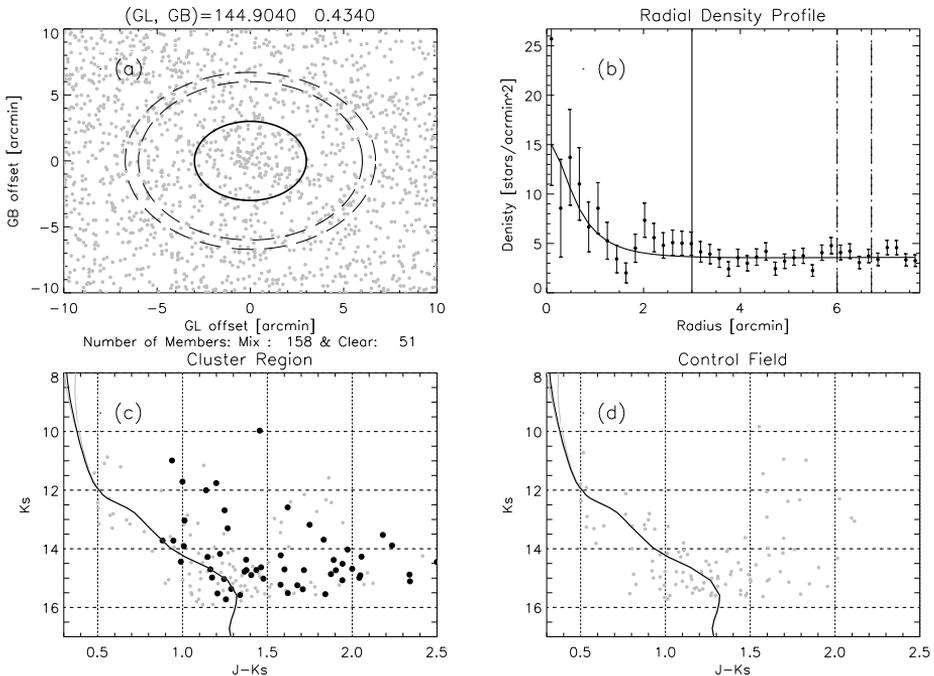}
\caption{\label{all} Analysis of HDG\,01 in Fig.~\ref{sample}. (a)~shows the spatial distribution of 
2MASS sources.  (b)~shows the radial density profile fit with a King's model (thin curve).  The thick 
line marks the apparent boundary of the cluster, $3'$, and the dashed lines delimit a field region twice 
the radius away, having the same sky area as the cluster region.  The boundaries of the cluster and field 
regions are shown in (a) by a thick solid circle, and thin dashed circles, respectively.  (c)~The 
color-magnitude diagram of HDG\,01, for all (in gray) 2MASS sources within the cluster region, i.e., 
with a radius $< 3'$, and those within $1.5'$ (in black) thus having a higher probability being 
member stars.  (d)~The same color-magnitude diagram for the field region, i.e., with a radius $> 6.0'$.  
The solid line is the isochrone of $\log(t)=7.0$~yr eye-fitted to the HDG\,01 members.  Dashed lines 
are $\log(t)=6.7$ and $\log(t)=7.5$~yr, respectively, for reference.  The distance modulus is 
$1.5\pm0.5$~kpc.}
\end{figure*}

\section{Characterization with 2MASS}
Figure~\ref{all} shows our analysis of HDG\,01 in Fig.~\ref{sample}.  The radial density profile suggested 
a field density of $4.05\pm1.07$~stars per square arcmin, and a density enhancement within $3'$.  The 
central density in the cluster region reaches $26.66\pm8.11$~stars per square arcmin, and there are a total 
of 51 candidate member stars within the $3'$ cluster radius. 

The age and distance of NDG\,01 are derived by fitting Padova 2MASS photometric isochrones with solar 
metallicity\footnote{http://stev.oapd.inaf.it/cmd}.  The best fit, as judged by eye, gives a distance 
of $1.5\pm0.5$~kpc.  At this distance, the cluster's apparent angular radius of $3'$ corresponds to a 
physical radius of 3.7~pc.  With no kinematic information, particularly for giant stars to constrain the
 isochrone, the age of HDG\,01 is very uncertain.  Due to some nebulous shape around this region, it should 
 have an young-age, less then 10~Myr.

\section{Summary}
We have developed a star counting algorithm and tested it on the 2MASS star catalog to search for candidate 
star clusters.  Within the Galactic latitude $|b|<50^{\circ}$, some 80\% overdensity regions are associated 
with known star clusters, or other objects (galaxies or clusters of galacies), and a total of 52 uncharted 
cluster candidates have been found.  Here we present one such candidate, HDG\,01 at 
$(\ell, b)=(144.9038, 0.4338)$, with nearly 80 member candidates with an angular radius of $3'$, and at a 
heliocentric distance of 1.8~kpc.  The full analysis will be presented elsewhere.  Our algorithm will be 
applied to other sky survey catalogs such as the Pan-STARRS, which reaches much fainter brightness limits 
(22--23 mag in r$'$) than those of 2MASS. We expect to enlarge the known OC sample to a much larger space 
volume in the solar neighborhood and down to significantly lower stellar mass than data currently available.  
In the third and fourth quadrants, in particular, where the extinction is relatively low, we will be able 
to explore the entire Milky Way disk, thereby providing a comprehensive sample of OCs to probe the formation 
and evolutionary history of the Galactic disks.

This work is supported by the grant NSC98-2917-I-008-103 of the National Science Council of Taiwan.

\end{document}